# Influence of Sensor Tilts on Bio-inspired Polarized Skylight Orientation Determination

LIANG Hua-ju[1], BAI Hong-yang[1*], ZHOU Tong[2], SHEN Kai[3]

(*[1] School of Energy and Power Engineering, Nanjing University of Science and Technology, Nanjing 210094, China*

*[2] School of Mechanical Engineering, Nanjing University of Science and Technology, Nanjing 210094, China*

*[3] School of Automation, Beijing Institute of Technology, Beijing 100081, China*)

*\* Corresponding Author, E-mail: hongyang@njust.edu.cn*

**Abstract**： Inspired by many insects, several polarized skylight orientation determination approaches have been proposed. However, almost all of these approaches always require polarization sensor pointing to the zenith of the sky dome. So, the influence of sensor tilts (not point to the sky zenith) on bio-inspired polarization orientation determination needs to be analyzed urgently. Aiming at this problem, a polarization compass simulation system is designed based upon solar position model, Rayleigh sky model, and hypothetical polarization imager. Then, the error characteristics of four typical orientation determination approaches are investigated in detail under only pitch tilt condition, only roll tilt condition, pitch and roll tilts condition respectively. Finally, simulation and field experiments all show that the orientation errors of four typical approaches are highly consistent when they are subjected to tilt interference, in addition, the errors are affected by not only the degree of inclination, but also the solar altitude angle and the relative position between the Sun and polarization sensor. The results of this paper can be used to estimate the orientation determination error caused by sensor tilts and correct this kind of error.

Supported by National Natural Science Foundation of China (61603189);

# 1 Introduction

Traditional navigation systems such as Inertial navigation system, Global Position System (GPS) and Geomagnetic Navigation System (GNS) play a key role in navigation for aircraft, robots, missiles, vehicles and so on. Inertial navigation system has many advantages, whereas the gyroscopes and accelerometers are usually prone to drifts and noises, which may cause errors to accumulate over time [1]. In addition, although GPS is a real-time and cheap locating system, GPS signal can be easily jammed due to the presence of disturbances [2, 3]. Besides, GNS is sensitive to electromagnetic interference [4]. With the fast development of human society, there is an urgent need to design a highly precise, autonomous, reliable, and robust navigation system.

Animals' navigation behavior provides us with a new idea about navigation [5, 6]. Desert ants rely on the predictable pattern of polarized light in the sky to find their way back home in hostile environments [7, 8]. Honey bees are able to detect the polarization of skylight to journey from hives [9, 10]. Birds may use skylight polarization patterns to calibrate magnetic compasses during longer-range migratory[11]. Dorsal rim area (DRA) of locusts' compound eyes is sensitive to the polarized skylight, thus can estimate its orientation [12, 13].

The reason why these animals can detect polarized light as a compass is that there is a polarization pattern in the sky [14, 15]. Unpolarized sunlight through the Earth's atmosphere produces the skylight polarization pattern [16-19]. Sunlight remains unpolarized until interacting with the atmospheric constituents, scattering sunlight causes a partial linear pattern of polarization in the sky, which can be well described by Rayleigh sky model [20-22].

Inspired by animals' polarization navigation behaviours, several orientation determination methods have been proposed based on Rayleigh sky model. Polarization orientation determination methods mainly include the following four typical approaches: Zenith approach, SM-ASM (solar meridian and anti-solar meridian) approach, Symmetry approach and Least-square approach. The heading angle is determined by measuring the angle of polarization (AOP) at the sky zenith, which is named Zenith approach [3, 16, 23-25]. The polarization E-vector along SM-ASM is consistently perpendicular to SM-ASM, so the heading angle can be calculated by extracting SM-ASM, which is named SM-ASM approach [26-29]. Because of the symmetry of the skylight polarization pattern, symmetry detection can be used to determine orientation, which is named Symmetry approach [30-32]. Polarization E-vector of Rayleigh sky model is consistently perpendicular to the solar vector, so the orientation can be determined by total least square method, which is then named Least-square approach [33-36]. However, all most of these heading determination approaches require the polarization sensor point toward the zenith of sky dome [37]: Zenith approach needs to directly capture the polarization information at the sky zenith; SM-ASM and Symmetry approaches require that reference direction of AOP is converted to the local meridian using the sky zenith as a reference point; Least-square approach requires a known sky zenith dependent coordinate system to accurately determine orientation.

In actual navigation, the carriers moving in three-dimensional space such as aerial vehicles, aircraft, and rockets will tilt. Even the carriers moving on the ground such as vehicles and multi-legged robots will tilt when the ground is uneven [38, 39]. So, the influence of polarization sensor tilts on bio-inspired polarized skylight heading determination urgently needs to be investigated and discussed in detail [15].

To summarize, this paper aims to make a profound study on the influence of sensor tilts for polarization orientation determination. Firstly, a polarization compass simulation system is designed. Secondly, based on this simulation system, numerical simulation experiments are carried out to investigate the influence of sensor tilt on the above four classical heading determination approaches. Finally, the results of field experiments are compared with digital simulation to further verify our conclusions.

# 2 Polarization Simulation System

To investigate the influence of sensor tilts on orientation determination, a polarization compass simulation system is designed, which is shown in Fig. 1.

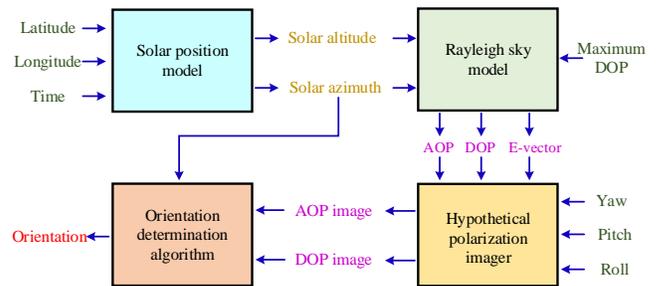

Fig.1 Polarization compass simulation system, where AOP represents the angle of polarization, DOP represents the degree of polarization, E-vector represents the polarization electric field vector. Green words indicate input, the orange words indicate input or process data, pink words indicate process data and red word indicates output.

In addition, Sun azimuth coordinate frame is constructed to better describe this system. As shown in Fig. 2, $ox_g y_g z_g$ is the East-North-Up (ENU) geography coordinate frame. The $y_a$ axis of Sun azimuth coordinate frame $ox_a y_a z_a$ is aligned with the solar azimuth, the $z_a$ axis points to the zenith, and the $x_a$ axis completes the right-handed coordinate frame. The Sun azimuth coordinate frame rotates around $z_a$ axis when the solar azimuth changes, and the direction of $y_a$ axis is always aligned with the direction of solar azimuth.

The rotation matrix from ENU coordinate to Sun azimuth coordinate is

$$C_g^a = \begin{bmatrix} \cos\varphi_{gS} & -\sin\varphi_{gS} & 0 \\ \sin\varphi_{gS} & \cos\varphi_{gS} & 0 \\ 0 & 0 & 1 \end{bmatrix} \quad (1)$$

where $\varphi_{gS}$ is the solar azimuth angle in the ENU coordinate.

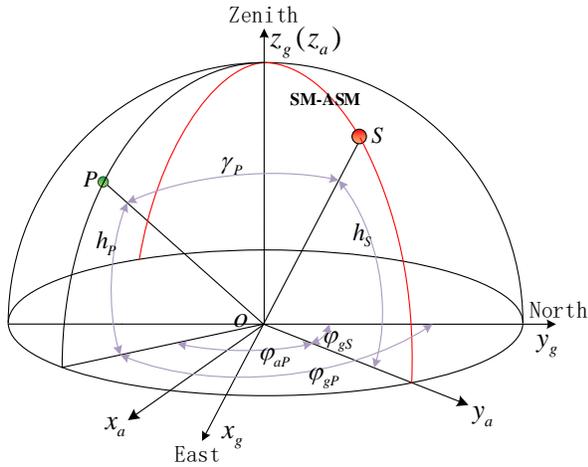

Fig.2 Coordinate frame: $ox_g y_g z_g$ is the East-North-Up (ENU) geography coordinate frame. $ox_a y_a z_a$ is the Sun azimuth coordinate frame. The red point $S$ represents the Sun. The green point $P$ represents the observation point. The red line SM-ASM represents the solar meridian and anti-solar meridian. $h_S$ is the solar altitude angle. $h_P$ is the altitude angle of $P$. $\gamma_P$ is the angle between $P$ and Sun. $\varphi_{gS}$ is the solar azimuth angle and $\varphi_{gP}$ is the azimuth angle of $P$ in ENU coordinate. $\varphi_{aP}$ is the azimuth angle of $P$ in Sun azimuth coordinate.

## 2.1 Solar Position Model

In this part, the solar position is calculated by the relevant formulas of astronomy [40-43] to facilitate the comparison between the simulation results and the field experiment results.

According to the relevant formula of astronomy, the position of the Sun can be solved through three angles, which are solar declination angle $\delta_S$, solar hour angle $T_S$ and latitude $L_O$ of observing site. By solving the spherical triangle $S-O-NP$ in Fig. 3, the solar altitude angle $h_S \in [0°, 90°]$ and solar azimuth angle $\varphi_{gS} \in [0°, 360°]$ in ENU coordinate frame are given by

$$\begin{cases} \sin h_S = \sin\delta_S \sin L_O + \cos\delta_S \cos L_O \cos T_S \\ \cos\varphi_{gS} = \dfrac{\sin\delta_S - \sin h_S \sin L_O}{\cos h_S \cos L_O} \end{cases} \quad (2)$$

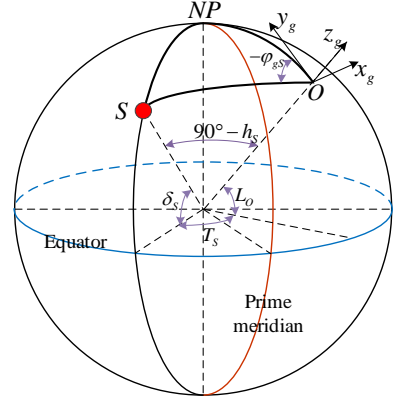

Fig.3 Celestial sphere, where red point $S$ represents the Sun, O represents the observing site, $NP$ represents the North Pole. $\delta_S$ is the solar declination angle, $T_S$ is the solar hour angle and $L_O$ is the latitude of observing site, $h_S$ is the solar altitude angle. $\varphi_{gS}$ is the solar azimuth angle in ENU coordinate.

By Solving the inverse trigonometric function of Eq. (2) and making quadrant judgment, then we have

$$h_S = \arcsin(\sin\delta_S \sin L_O + \cos\delta_S \cos L_O \cos T_S) \quad (3)$$

$$\varphi_{gS} = \begin{cases} \arccos\left(\dfrac{\sin\delta_S - \sin h_S \sin L_O}{\cos h_S \cos L_O}\right) & T_S < 0 \\ 360° - \arccos\left(\dfrac{\sin\delta_S - \sin h_S \sin L_O}{\cos h_S \cos L_O}\right) & T_S > 0 \end{cases} \quad (4)$$

where the formula of solar declination angle $\delta_S$ for a particular year 1985 is given by

$$\delta_S = 0.3723 + 23.2567\sin\sigma_S + 0.1149\sin 2\sigma_S - 0.1712\sin 3\sigma_S - 0.758\cos\sigma_S + 0.3656\cos 2\sigma_S + 0.0201\cos 3\sigma_S \quad (5)$$

where day angle $\sigma_S = 2\pi(D - D_0)/365.2422$, $D$ is day of year, and the spring-equinox time $D_0$ expressed in days from the particular year 1985 is

$$D_0 = 79.6764 + 0.2422 \times (Y - 1985) - \text{INT}\left[(Y - 1985)/4\right] \quad (6)$$

where $Y$ is the year and INT represents rounding down.

Calculation progress of the solar hour angle $T_S$ is shown below, where the local standard time $S_d$ of observing site can be calculated by

$$S_d = S_O + \left(F_O - 4(120° - Lon_O)\right)/60 \quad (7)$$

In Eq. (7), for observing site $O$, $S_O$ and $F_O$ are the hour and minute of Beijing time, $Lon_O$ is the longitude of observing site $O$.

Then, time error $E_t$ is given by

$$E_t = 0.0028 - 1.9857\sin\sigma_S + 9.9059\sin 2\sigma_S - 7.0924\cos\sigma_S - 0.6882\cos 2\sigma_S \quad (8)$$

After that, $S_d$ is corrected by $E_t$ to obtain solar time $S_t$.

$$S_t = S_d + E_t/60 \quad (9)$$

Finally, the solar hour angle $T_S$ is given by

$$T_S = (S_t - 12) \times 15° \quad (10)$$

In short, through the above formulae, the solar azimuth angle and solar altitude angle can be finally calculated and obtained.

## 2.2 Rayleigh Sky Model

Rayleigh sky model predicts the sky polarization properties degree of polarization (DOP) [3] as

$$DOP = DOP_{max} \frac{\sin^2\gamma_P}{1+\cos^2\gamma_P} \quad (11)$$

where $\gamma_P$ is the angle between observation point $P$ and Sun, which is named scattering angle. $DOP_{max}$ is the maximum detected DOP in the sky and $DOP_{max} = 1$ for an ideal sky.

Rayleigh sky model predicts the sky polarization properties AOP [26] as

$$AOP = \arctan\frac{\sin h_S \cos h_P - \cos h_S \sin h_P \cos(\varphi_{gS} - \varphi_{gP})}{\sin(\varphi_{gS} - \varphi_{gP})\cos h_S} \quad (12)$$

where $h_P$ is the altitude angle of observation point $P$, $\varphi_{gP}$ is the azimuth angle of $P$ in ENU coordinate, and $AOP \in (-90, 90)$. According to trigonometric functions,

$$\begin{cases} \dfrac{\sin AOP}{\cos AOP} = \tan AOP \\ \sin^2 AOP + \cos^2 AOP = 1 \end{cases} \quad (13)$$

Then, the $\sin AOP$ and $\cos AOP$ are given by

$$\begin{cases} \sin AOP = \dfrac{\sin h_S \cos h_P - \cos h_S \sin h_P \cos(\varphi_{gS} - \varphi_{gP})}{\sin\gamma_P} \\ \cos AOP = \dfrac{\sin(\varphi_{gS} - \varphi_{gP})}{\sin\gamma_P}\cos h_S \end{cases}$$
(14)

Then, the sky polarization E-vector in ENU coordinate predicted by Rayleigh sky model is given by

$$\vec{E}_{gP} = \vec{V}_{gP}\cos AOP + \vec{H}_{gP}\sin AOP \quad (15)$$

where $\vec{E}_{gP}$ represents the polarization E-vector of observation point $P$ in ENU coordinate predicted, $\vec{V}_{gP}$ represents the tangent direction of local meridian, and $\vec{H}_{gP}$ represents the vector, which is perpendicular to $\vec{V}_{gP}$ and parallel to plane $ox_gy_g$. $\vec{V}_{gP}\cos AOP$ represents the projection of polarization E-vector on $\vec{V}_{gP}$, and $\vec{H}_{gP}\sin AOP$ represents the projection of polarization E-vector on $\vec{H}_{gP}$.

$$\vec{V}_{gP} = (-\sin h_P \sin\varphi_{gP}, -\sin h_P \cos\varphi_{gP}, \cos h_P)^T \quad (16)$$

$$\vec{H}_{gP} = (-\cos\varphi_{gP}, \sin\varphi_{gP}, 0)^T \quad (17)$$

The superscript $T$ represents matrix or vector transpose. Substituting (14) into (15), $\vec{E}_{gP}$ in ENU coordinate is given by

$$\vec{E}_{gP} = \begin{bmatrix} (\cos\varphi_{gS}\sin h_P \cos h_S - \cos\varphi_{gP}\cos h_P \sin h_S)/\sin\gamma_P \\ (-\sin\varphi_{gS}\sin h_P \cos h_S + \sin\varphi_{gP}\cos h_P \sin h_S)/\sin\gamma_P \\ \sin(\varphi_{gS} - \varphi_{gP})\cos h_S \cos h_P /\sin\gamma_P \end{bmatrix}$$
(18)

In short, the polarization E-vector $\vec{E}_{aP}$ in Sun azimuth coordinate can be given by

$$\vec{E}_{aP} = C_g^a \vec{E}_{gP} \quad (19)$$

## 2.3 Hypothetical Polarization Imager

In order to construct a comprehensive and perfect simulation system, not only skylight polarization model, but also polarization imaging sensor need to be constructed [37]. In this section, a hypothetical polarization imager is designed and described in detail.

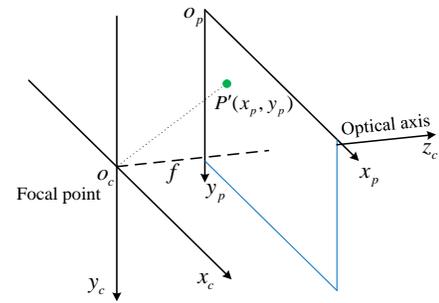

Fig.4 Camera coordinate frame $o_c x_c y_c z_c$ and pixel coordinate frame $o_p x_p y_p$, where $f$ is focal length.

To construct hypothetical polarization imager, camera coordinate and pixel coordinate frames are established, as shown in Fig. 4. The $z_c$ axis of camera coordinate frame $o_c x_c y_c z_c$ is aligned with the optical axis of imager, the $x_c$ and $y_c$ axes of camera coordinate frame $o_c x_c y_c z_c$ are aligned with the column and row directions of image, respectively. The $x_p$ and $y_p$ axes of pixel coordinate frame $o_p x_p y_p$ are aligned with the column and row directions of image respectively and the unit of this coordinate is pixel.

In the camera coordinate frame, the vector of pixel

$P'(x_{pP'}, y_{pP'})$ is

$$\overrightarrow{V_{cP'}} = \left( D_x(x_{pP'} - \frac{\eta_x+1}{2}), D_y(y_{pP'} - \frac{\eta_y+1}{2}), f \right)^T \quad (20)$$

where $D_x$ and $D_y$ represent the column and row pixel size, respectively, $\eta_x$ and $\eta_y$ indicate the polarization image has $\eta_x \times \eta_y$ pixels, and $f$ is the focal length of pixel-based polarization camera that we used.

Suppose three Euler angles of polarization imager are given, then the rotation matrix from camera coordinate to Sun azimuth coordinate can be described as
$C_c^a =$

$$\begin{bmatrix} \cos\beta\cos\psi + \sin\beta\sin\alpha\sin\psi & \cos\alpha\sin\psi & \sin\beta\cos\psi - \cos\beta\sin\alpha\sin\psi \\ -\cos\beta\sin\psi + \sin\beta\sin\alpha\cos\psi & \cos\alpha\cos\psi & -\sin\beta\sin\psi - \cos\beta\sin\alpha\cos\psi \\ -\sin\beta\cos\alpha & \sin\alpha & \cos\beta\cos\alpha \end{bmatrix}$$
(21)

where $\psi$, $\alpha$ and $\beta$ represent yaw, pitch and roll angle, respectively.

Then, the shooting direction of pixel $P'$ in Sun azimuth coordinate is

$$\overrightarrow{V_{aP'}} = C_c^a \overrightarrow{V_{cP'}} \quad (22)$$

Azimuth angle $\varphi_{aP'}$ of the shooting direction of pixel $P'$ in Sun azimuth coordinate is

$$\varphi_{aP'} = \arctan\left( \frac{\overrightarrow{V_{aP'}}(1,1)}{\overrightarrow{V_{aP'}}(2,1)} \right) \quad (23)$$

Altitude angle $h_{P'}$ of the shooting direction of pixel $P'$ is

$$h_{P'} = \arcsin\left( \frac{\overrightarrow{V_{aP'}}(3,1)}{|\overrightarrow{V_{aP'}}|} \right) \quad (24)$$

where $\overrightarrow{V_{aP'}}(1,1)$, $\overrightarrow{V_{aP'}}(2,1)$ and $\overrightarrow{V_{aP'}}(3,1)$ are the components of $\overrightarrow{V_{aP'}}$, $|\overrightarrow{V_{aP'}}|$ is the mode of $\overrightarrow{V_{aP'}}$. Then, the scattering angle $\gamma_{P'}$ of pixel $P'$ is given by

$$\gamma_{P'} = \arccos(\sin h_{P'} \sin h_S + \cos h_{P'} \cos h_S \cos(\varphi_{aP'})) \quad (25)$$

Substituting Eq. (25) into Eq. (11), the DOP of pixel $P'$ can be obtained and Fig. 5(a) shows a hypothetical DOP image.

The azimuth angle $\varphi_{gP'}$ of the shooting direction of pixel $P'$ in ENU coordinate is

$$\varphi_{gP'} = \varphi_{aP'} - \varphi_{gS} \quad (26)$$

So, substituting Eq. (24), Eq. (26) and Eq. (18) into Eq. (19), the polarization E-vector $\overrightarrow{E_{aP'}}$ of pixel $P'$ in Sun azimuth coordinate frame can be obtained. Thus, the polarization E-vector $\overrightarrow{E_{cP'}}$ of $P'$ in camera coordinate frame can be given by

$$\overrightarrow{E_{cP'}} = C_a^c \overrightarrow{E_{aP'}} \quad (27)$$

where $C_a^c$ is the transpose of $C_c^a$, which represents the rotation matrix from Sun azimuth coordinate to camera coordinate. As the AOP reference direction is aligned with $y_b$ axis and the shooting direction of the hypothetical polarization imager is aligned with $z_c$ axis, then AOP can be given by

$$AOP = \arctan\frac{\overrightarrow{E_{cP'}}(1,1)}{\overrightarrow{E_{cP'}}(2,1)} \quad (28)$$

where $\overrightarrow{E_{cP'}}(1,1)$, $\overrightarrow{E_{cP'}}(2,1)$ are the components of $\overrightarrow{E_{cP'}}$, and a hypothetical AOP image is shown in Fig. 5(b).

For the classical four typical orientation determination algorithms described in Section 1, Zenith approach and Least-square approach can directly use AOP for orientation determination. However, for SM-ASM approach and Symmetry approach, further transformation of AOP is required, the reference direction of AOP needs to be converted to the local meridian, the AOP whose reference direction is local meridian can be defined as AOPLM.

$$AOPLM = AOP - \xi \quad (29)$$

where $\xi$ is the angle between $y_c$ axis and local meridian. When polarization imager points to the sky zenith, we have

$$\xi = \arctan\left( \frac{(x_{pP'} - \frac{\eta_x+1}{2})}{(y_{pP'} - \frac{\eta_y+1}{2})} \right) \quad (30)$$

And a hypothetical AOPLM image is obtained and shown in Fig. 5(c).

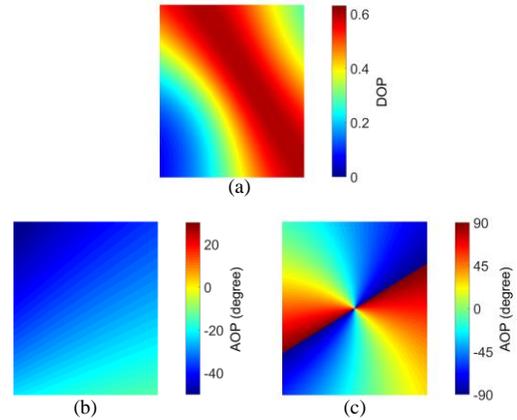

Fig.5 Hypothetical polarization images: (a) DOP image; (b) AOP image; (c) AOPLM image.

## 3 Simulation

In order to investigate the influence of sensor tilts on orientation determination, we have carried out a lot of simulation experiments for four classical polarization orientation determination algorithms: Zenith approach,

SM-ASM approach, Symmetry approach and Least-square approach, as shown in Fig. 6. According to Rayleigh sky model, the polarization E-vector at the sky zenith is perpendicular to the solar azimuth, so, Zenith approach determines the heading angle by measuring the AOP at the sky zenith [3, 16, 23-25]. The polarization E-vector along SM-ASM is consistently perpendicular to SM-ASM, so SM-ASM approach calculates the heading angle by extracting SM-ASM [26-29]. According to the symmetry of the skylight polarization pattern, Symmetry approach determines orientation by symmetry detection [30-32]. Polarization E-vector of Rayleigh sky model is consistently perpendicular to the solar vector, so, Least-square approach determines orientation by total least square of Polarization E-vectors [33-35].

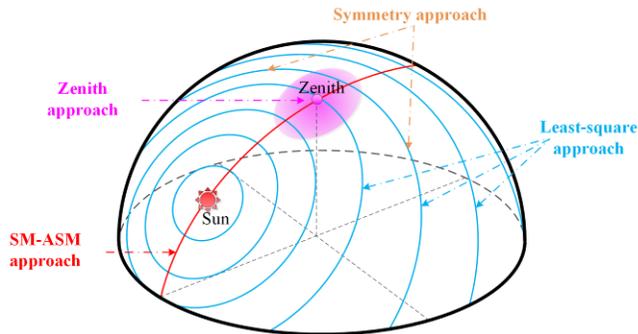

Fig.6 Rayleigh sky model and four typical orientation determination approaches, where the red point represents the Sun, the pink point represents the sky zenith, the red line represents solar meridian and anti-solar meridian (SM-ASM), the blue lines represent the polarization electric field vectors (E-vector).

**Table 1** Simulation Parameters.

| Symbol | Value | Units | Description |
| --- | --- | --- | --- |
| $DOP_{max}$ | 1 | / | Maximum DOP in the sky |
| $D_x$ | 3.45 | μm | Pixel size in column direction |
| $D_y$ | 3.45 | μm | Pixel size in row direction |
| $\eta_x$ | 2048 | pixel | Number of pixels in column direction |
| $\eta_y$ | 2448 | pixel | Number of pixels in row direction |
| $f$ | 4 | mm | Focal length of polarization imager |

Considering that the polarization imager needs to capture the skylight polarization pattern, the imager field of view should always be above the horizon, and the interference of buildings and obstacles should be eliminated. In our simulation and experiment, the imager angle of view is 108°, therefore, we set pitch and roll angles to be $|\alpha|+|\beta| \leq 30°$.

The tilt state of the sensor in practice can be divided into three situations:
  1) Only pitch tilt condition
  2) Only roll tilt condition
  3) Pitch and roll tilts condition
  And the parameters of simulation are shown in Table 1.

## 3.1 Only Pitch Tilt

This part discusses the error characteristics when the tilt is only the pitch angle with roll angle set to zero. By using the polarization compass simulation system in Section 2, more than $1.4 \times 10^5$ sets of simulation experiments were carried out, and the orientation errors of four typical approaches were obtained in the range of solar altitude angle $h_s \in [0°, 50°]$, yaw angle $\psi \in [-180°, 180°]$, pitch angle $\alpha \in [-30°, 30°]$ and roll angle $\beta = 0°$. And the results under only pitch tilt are shown in Fig. 7.

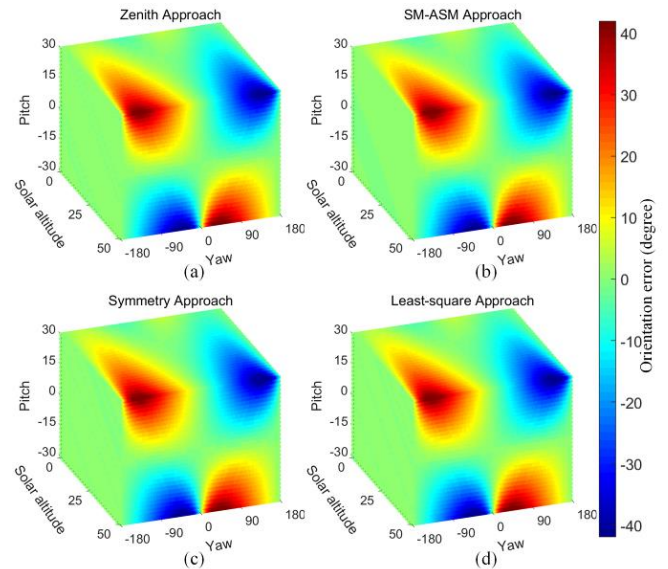

Fig.7 Orientation errors of four typical polarization orientation determination approaches under only pitch tilt: (a) Orientation error of Zenith approach; (b) Orientation error of SM-ASM approach; (c) Orientation error of Symmetry approach; (d) Orientation error of Least-square approach. The unit of the three axes is degree.

It can be observed in Fig. 7, when only pitch tilt exists, the variation trend of orientation error of the four typical approaches is the same. And there are three following similarities:

(Ⅰ). When the pitch angle is 0°, errors of the four approaches are all close to zero. With the increase in pitch angle, the errors of the four approaches all have a tend to increase.

(Ⅱ). When the solar altitude angle is 0°, the errors of the four approaches are always close to 0°. When the pitch angle is not 0°, the errors of the four approaches tend to increase with the increase in the solar altitude angle.

(Ⅲ). The errors of the four approaches are all symmetric with respect to the plane $\psi = 0°$ and $\psi = 180°(-180°)$. And when the yaw angle is 0° or $180°(-180°)$, no matter what the pitch and solar altitude angles are, errors of the four typical approaches are always close to zero. In addition, there are the following trends: Errors are close to zero at $\psi=-180°$, and with the yaw angle increasing gradually, errors increase gradually and reach a maximum; Then, errors decrease and are close to zero at $\psi=0°$; After that, with the yaw angle increasing gradually, orientation calculating errors increase gradually and reach a maximum; Finally, the orientation errors decrease and are close to zero at $\psi=180°$.

The following is a detailed analysis of the reasons for the above three similarities:

For (Ⅰ), when the pitch angle $\alpha=0°$, there is no influence of sensor tilt, which means, under ideal conditions, all these four typical approaches can effectively determine the orientation. When pitch tilt increases, the tilt interference increases, which leads to the increase in orientation errors.

For (Ⅱ), the four approaches essentially use the solar azimuth information to determine orientation. When the solar altitude angle increases, the component of the solar vector projected on the plane $ox_a y_a (ox_g y_g)$ decreases, so the stability and reliability of solar azimuth are weakened, which leads to the increase in orientation errors.

For (Ⅲ), The errors of the four approaches are all symmetric with respect to the planes $\psi=0°$ and $\psi=180°(-180°)$. This manifests the symmetry of skylight polarization pattern with respect to the SM-ASM. When the yaw angle is $0°$ or $180°(-180°)$ and only pitch tilt, the direction of polarization imager's optical axis always points to SM-ASM which is parallel to $y_c$ axis, thus results in that when the yaw angle is $0°$ or $180°(-180°)$, no matter what the pitch and solar altitude angles are, the errors of the four approaches are always close to zero.

In addition, to further compare these four approaches, we have drawn groups of simulation results of the four approaches on a graph, as shown in Fig. 8 and 9. It can be found that the variation trend of the four error curves is exactly the same, and the four curves almost coincide. Furthermore, under the same condition, the error difference of these four approaches is always less than 0.66°. Therefore, when there is only pitch tilt interference, it can be concluded that the error characteristics of the four approaches are consistent and the orientation errors of the four approaches are almost the same. Moreover, similarity (Ⅱ) can be clearly seen from Fig. 8 and similarities (Ⅰ) and (Ⅲ) can be partially reflected in Fig. 9.

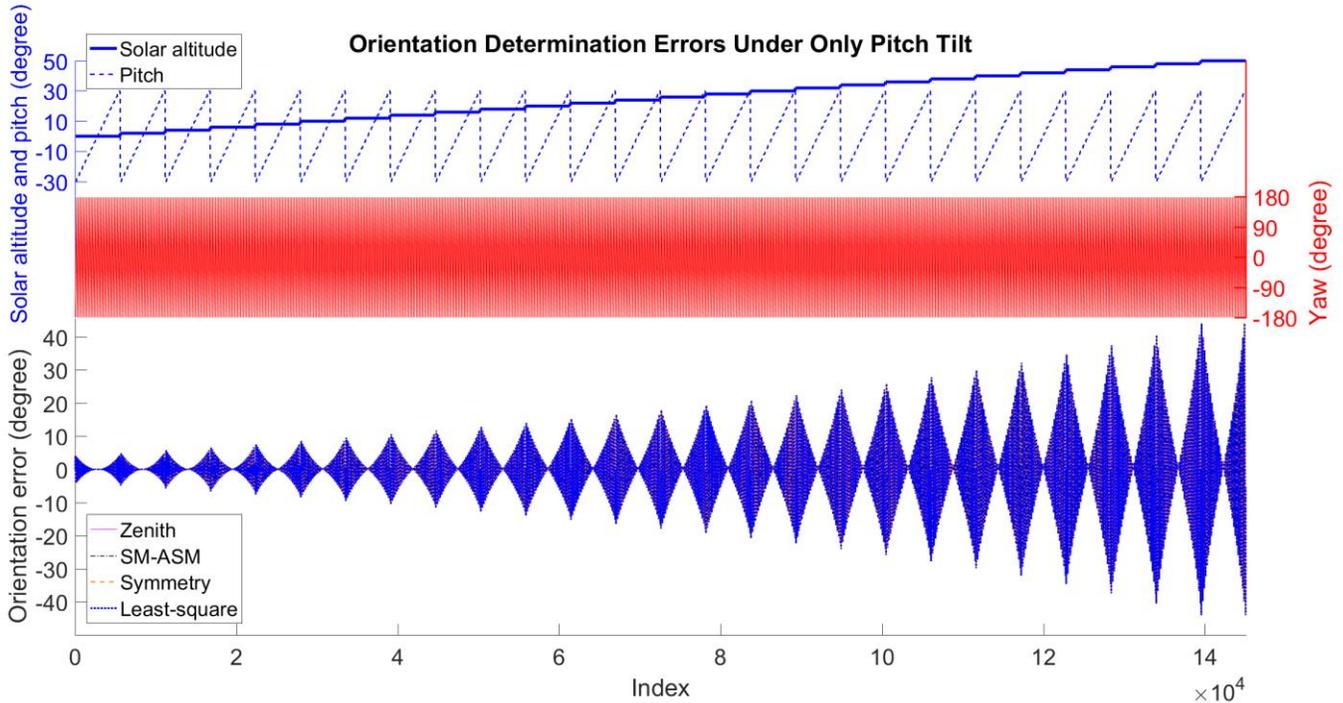

Fig.8 Simulation orientation error curves of four typical polarization orientation determination approaches under only pitch tilt situation.

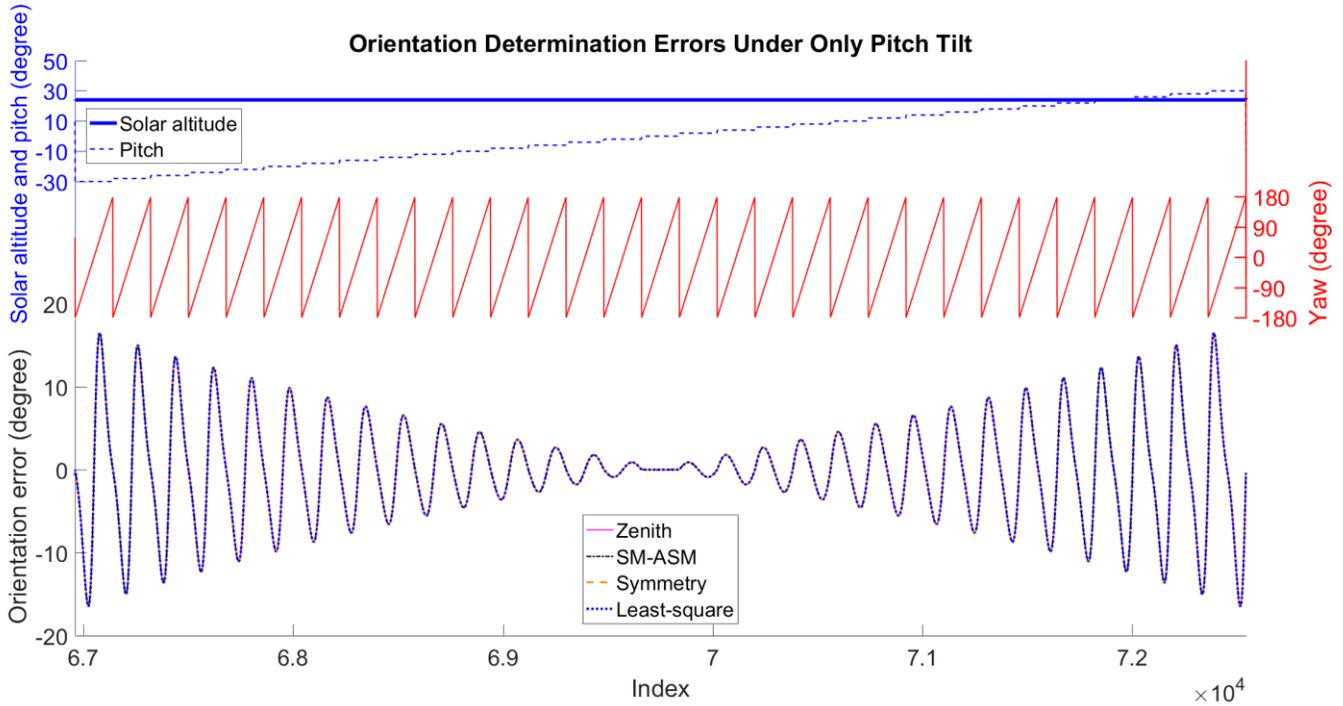

Fig.9 Simulation orientation error curves of four typical approaches under only pitch tilt situation, which is an enlarged view of Fig. 8.

### 3.2 Only Roll Tilt

This part discusses the error characteristics when it has only the roll angle tilt situation, with pitch angle set to be zero. The orientation errors of four typical approaches were obtained in the range of solar altitude angle $h_s \in [0°, 50°]$, yaw angle $\psi \in [-180°, 180°]$, roll angle $\beta \in [-30°, 30°]$ and pitch angle $\alpha = 0°$. The results under only roll tilt are shown in Fig. 10, where the yaw range is $[-180°, 180°]$. For ease of observation and comparison, the range of yaw is converted to $[-90°, 270°]$, as shown in Fig. 11. By comparing Fig. 11 and Fig. 7, it can be seen that the two sets of graphs have exactly the same shape, with the only difference being the range of yaw angle. Therefore, the error characteristics of only roll tilt are very similar to that of only pitch tilt. The first two error similarities are the same as described in Section 3.1, the only difference is the third one.

For only roll tilt, errors of the four approaches are all symmetric with respect to the plane $\psi = -90°(270°)$ and $\psi = 90°$. And when the yaw angle is $90°$ or $-90°(270°)$, no matter what the roll and solar altitude angles are, the errors of the four typical approaches are always close to zero. In addition, there are the following trends: The errors are close to zero at $\psi = -90°$, and with the yaw angle increasing gradually, the errors increase gradually and reach a maximum; Then, the errors decrease and are close to zero at $\psi = 90°$; After that, with the yaw angle increasing gradually, the errors increase gradually and reach a maximum; Finally, the errors decrease and are close to zero at $\psi = 270°$.

In short, compared with the only pitch tilt case, the result of the only roll tilt case has a $90°$ shift in the yaw direction. The reason for this phenomenon is explained in detail below.

As shown in Figure 5, the direction of polarization imager's optical axis is $(0,0,1)^T$. According to Eq. (22), the optical axis direction in Sun azimuth coordinate is given by

$$\vec{V_{af}} = C_c^a \begin{pmatrix} 0 \\ 0 \\ 1 \end{pmatrix} = \begin{pmatrix} \sin\beta\cos\psi - \cos\beta\sin\alpha\sin\psi \\ -\sin\beta\sin\psi - \cos\beta\sin\alpha\cos\psi \\ \cos\beta\cos\alpha \end{pmatrix} \quad (31)$$

Assume two sets of attitude $(\psi_1, \alpha_1, \beta_1)$ and $(\psi_2, \alpha_2, \beta_2)$, which satisfy $\psi_2 = \psi_1 + 90°$, $\beta_2 = \alpha_1$, $\beta_1 = 0°$ and $\alpha_2 = 0°$.

$$\vec{V_{af1}} = \begin{pmatrix} \sin\beta_1\cos\psi_1 - \cos\beta_1\sin\alpha_1\sin\psi_1 \\ -\sin\beta_1\sin\psi_1 - \cos\beta_1\sin\alpha_1\cos\psi_1 \\ \cos\beta_1\cos\alpha_1 \end{pmatrix} =$$

$$\begin{pmatrix} -\sin\alpha_1\sin\psi_1 \\ -\sin\alpha_1\cos\psi_1 \\ \cos\alpha_1 \end{pmatrix} = \begin{pmatrix} \sin\beta_2\cos\psi_2 \\ -\sin\beta_2\sin\psi_2 \\ \cos\beta_2 \end{pmatrix} =$$

$$\left.\begin{pmatrix} \sin\beta_2 \cos\psi_2 - \cos\beta_2 \sin\alpha_2 \sin\psi_2 \\ -\sin\beta_2 \sin\psi_2 - \cos\beta_2 \sin\alpha_2 \cos\psi_2 \\ \cos\beta_2 \cos\alpha_2 \end{pmatrix}\right) = \overrightarrow{V_{af2}} \quad (32)$$

where $\overrightarrow{V_{af1}}$ and $\overrightarrow{V_{af2}}$ are the optical axis directions at $(\psi_1,\alpha_1,\beta_1)$ and $(\psi_2,\alpha_2,\beta_2)$ in Sun azimuth coordinate system, respectively. $\overrightarrow{V_{af1}} = \overrightarrow{V_{af2}}$ shows that the image's optical axis directions at $(\psi_1,\alpha_1,\beta_1)$ and $(\psi_2,\alpha_2,\beta_2)$ are exactly the same, so the polarization information collected by the polarization imager at $(\psi_1,\alpha_1,\beta_1)$ and $(\psi_2,\alpha_2,\beta_2)$ corresponds to almost the same area of the sky. This is the reason that the results of only roll tilt have 90° shift in the yaw direction compared with that of only pitch tilt.

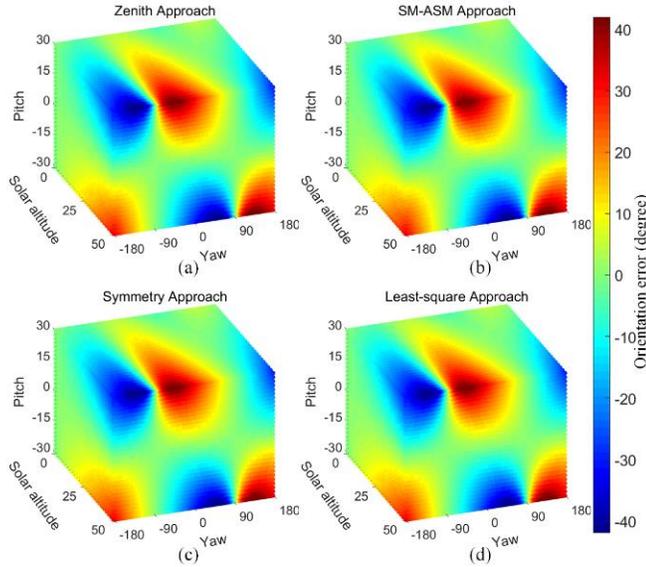

Fig.10 Orientation errors of four typical polarization orientation determination approaches under only roll tilt, where $\psi \in [-180°,180°]$: (a) Orientation error of Zenith approach; (b) Orientation error of SM-ASM approach; (c) Orientation error of Symmetry approach; (d) Orientation error of Least-square approach. The unit of the three axes is degree.

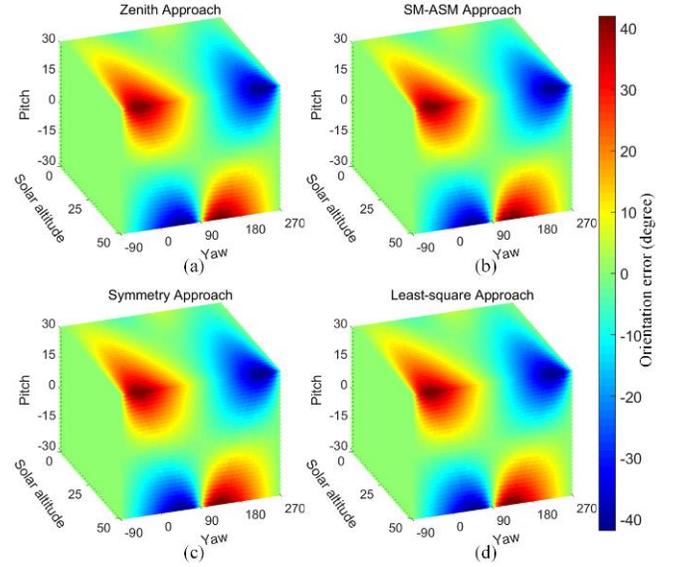

Fig.11 Orientation errors of four typical polarization orientation determination approaches under only roll tilt, where $\psi \in [-90°,270°]$: (a) Orientation error of Zenith approach; (b) Orientation error of SM-ASM approach; (c) Orientation error of Symmetry approach; (d) Orientation error of Least-square approach. The unit of the three axes is degree.

To further compare these four approaches under only roll tilt, we have drawn groups of simulation results of the four approaches on a graph, as shown in Fig. 12 and 13. It can be found that the variation trend of the four error curves is exactly the same, and the four curves almost coincide. Furthermore, as shown in Fig. 13, the error curve of only pitch tilt is drawn to compare with that of only roll tilt. It can be clearly seen, the results of only roll tilt have 90° shift in the yaw direction compared with that of only pitch tilt. Therefore, the properties of only roll tilt can be referred from that of only pitch tilt, which will not be repeated here.

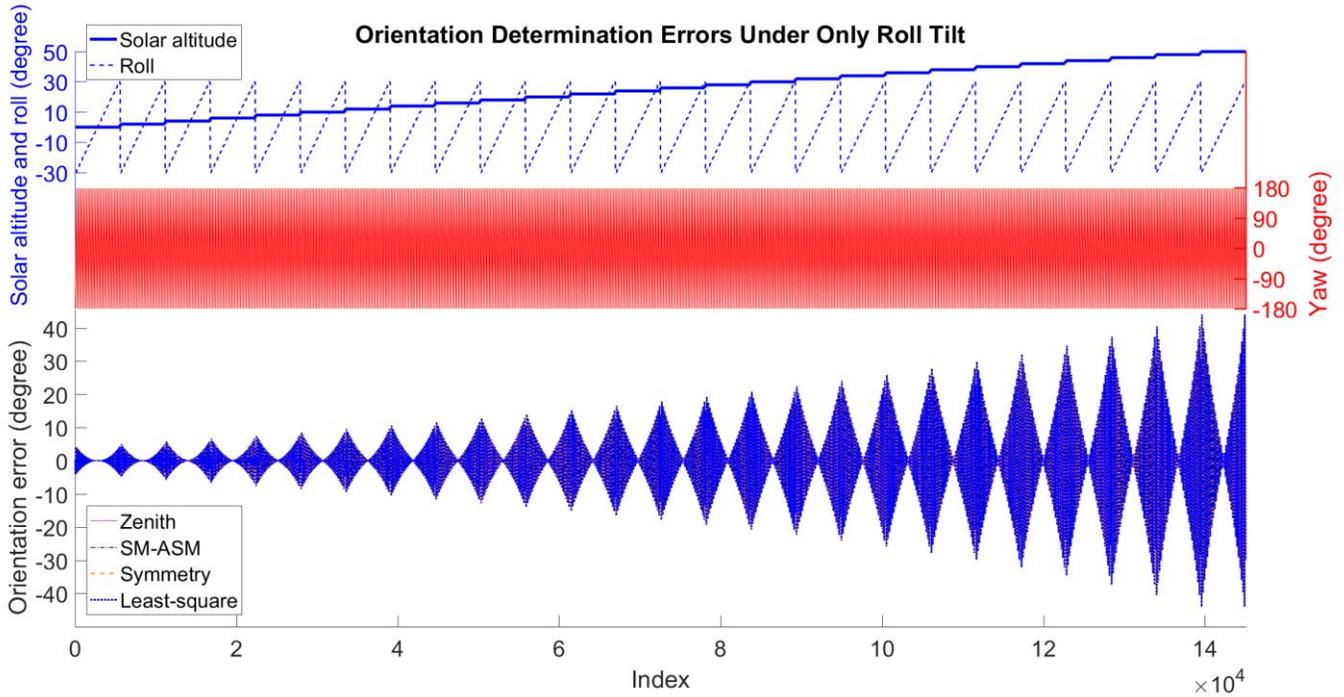

Fig.12 Simulation orientation error curves of four typical polarization orientation determination approaches under only roll tilt situation.

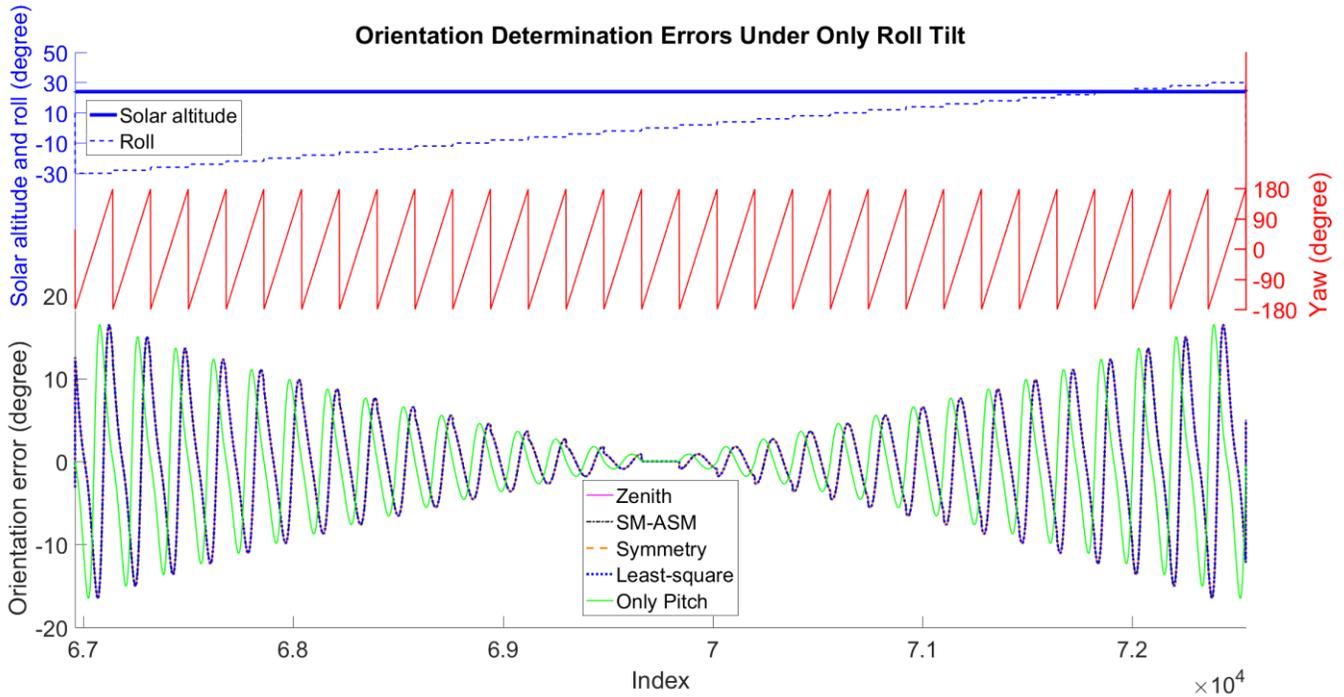

Fig.13 Simulation orientation error curves of four typical approaches under only roll tilt situation, which is an enlarged view of Fig. 12. Only Pitch is the error curve, when pitch angle equals roll angle under only pitch tilt situation.

### 3.3 Pitch and Roll Tilts

In addition, the error characteristics under pitch and roll tilts are discussed. The orientation errors of four typical approaches were obtained in the range of solar altitude angle $h_s \in [0°, 50°]$, yaw angle $\psi \in [-180°, 180°]$, pitch angle $\alpha \in [-30°, 30°]$, roll angle $\beta \in [-30°, 30°]$ and $|\alpha| + |\beta| \leq 30°$. The results of Zenith approach under pitch and roll tilts are shown in Fig. 14.

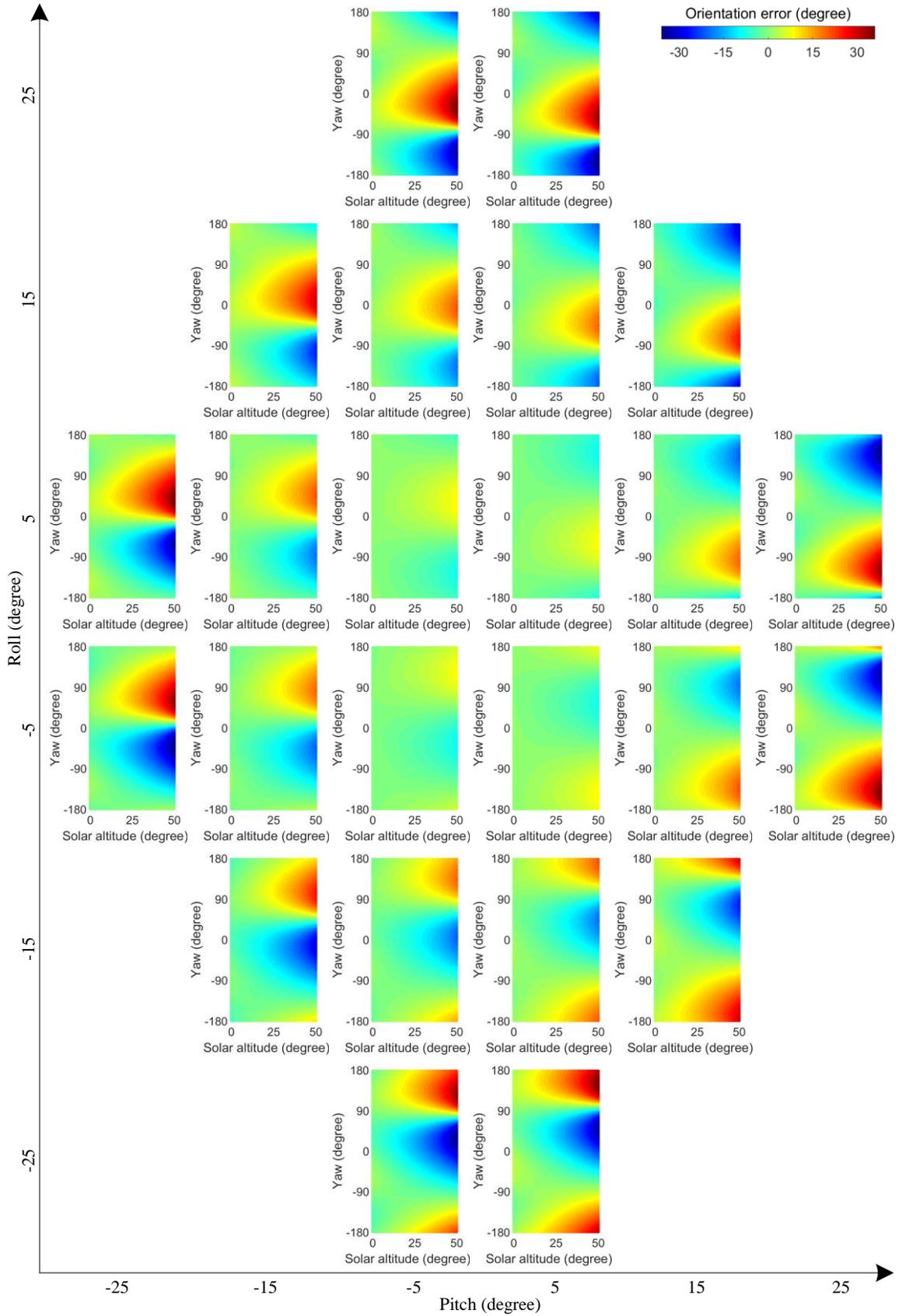

Fig.14 Simulation orientation error of Zenith approach under pitch and roll tilts.

It can be observed in Fig. 14, under pitch and roll tilts, the error of Zenith approach has the following characteristics:

(Ⅰ). With the increase in pitch and roll angles, the error of Zenith approach has a tendency to increase.

(Ⅱ). When the solar altitude angle is 0°, orientation error of Zenith approach is close to 0°. When the pitch and roll angles are not 0°, the error of Zenith approach tends to increase with the increase in the solar altitude angle.

(Ⅲ). When the other conditions are the same, yaw angle difference also affects orientation errors obviously.

The following will be a detailed analysis of the reasons for the above three characteristics. The reasons are identical to that mentioned in Section 3.1.

For (Ⅰ), when the pitch and roll tilts increase, the tilt interference increases, which leads to the increase in orientation errors.

For (Ⅱ), the Zenith approach essentially uses the solar azimuth information to determine orientation. When the solar altitude angle increases, the component of the solar vector projected on the plane $ox_a y_a (ox_g y_g)$ decreases, so the stability and reliability of solar azimuth are weakened, which leads to the increase in orientation error.

For (Ⅲ), with different yaw angles, the relative position between the Sun and polarization sensor is different, resulting in different orientation errors when the sensor tilts.

Under pitch and roll tilts, the error difference between the four approaches is always less than 0.77°. So, the error characteristics of the other three approaches are consistent with that of Zenith approach.

### 3.4 Application values

As mentioned above, the influence of sensor tilts has been discussed in detail. The analysis of this problem has two important application values: 1. Given the allowable error range of orientation, the allowable range of corresponding pitch and roll angles can be obtained. 2. When the pitch angle and roll angle are given, the error caused by sensor tilts can be corrected.

The allowable range of pitch and roll angle is illustrated by taking the allowable maximum error of orientation as an example. Suppose that the maximum allowable error of orientation is *ME*. It is important to note that, for any yaw angle, the maximum error of orientation is affected not only by pitch and roll angles, but also by solar altitude angle. When the pitch and roll angles are 0 degree, the orientation determination error caused by sensor tilts is 0 degree. When the pitch or roll angle increases, the maximum error of orientation determination increases. When the solar altitude angle is 0 degree, the orientation determination error is almost not affected by sensor tilt. When the solar altitude angle increases, the maximum error of orientation determination increases. Above all, according to the simulation results, the allowable range of pitch and roll angles can be roughly estimated by

$$h_s \sqrt{\alpha^2 + \beta^2} \leq ME^2 \quad (33)$$

Moreover, the results of this paper can also mitigate such orientation determination errors. The influence of sensor tilts is a kind of system error, and one way to eliminate this kind of error is to fully analysis the characteristics of the error. So, given pitch angle and roll angle, the error caused by tilts can be obtained based on the results of this paper. After that, the orientation can be calibrated by subtracting this error. In our field experiment, we mitigated such orientation determination errors as shown in Section 4.

## 4 Field Experiment

To further verify the simulation results, field experiments were carried to investigate the influence of polarization sensor tilts on orientation determination. In addition, the results of field experiments are compared with that of simulation.

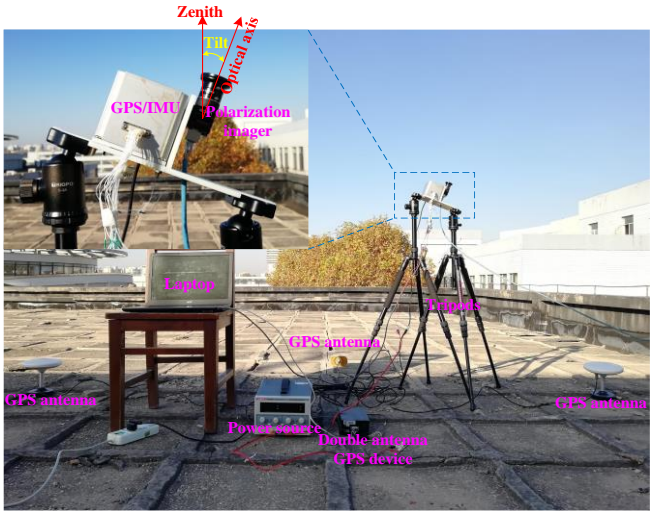

Fig.15 Polarization orientation determination experiment platform.

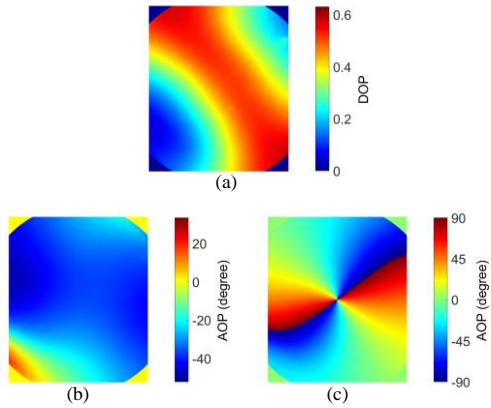

Fig.16 Actual polarization images: (a) DOP image; (b) AOP image; (c) AOPLM image. The four corners of these polarization images cannot be properly imaged due to the short focal lens adopted by the imager, so they are not used in polarized skylight navigation and are set to 0.

Our experiment platform is shown in Fig. 15. Two tripods are equipped with a Sony IMX250MZR polarization imager and a GPS/IMU (Global Position System/Inertial Measurement Units) integrated navigation system. The parameters of the actual polarization imager are also consistent with that of hypothetical polarization imager, as shown in Table 1. The GPS/IMU integrated navigation system is used to determine pitch and roll angles of polarization imager. The true North is determined by a double antenna GPS device as a benchmark (orientation resolution is 0.1° with a 2 m baseline). Field experiments were performed in Nanjing, China, on the roof of our laboratory (32°01′36.4″ N, 118°51′11.9″ E), from 15 November to 19 November 2019, meteorological conditions were stable. Fig. 16 shows a set of actual polarization images.

Aiming at the influence of sensor tilt, field experiments were carried out, which include only pitch tilt condition, only roll tilt condition, pitch and roll tilts condition. The experimental results are shown in Fig. 17, 18 and 19. Note that the green curves in Fig. 17, 18 and 19 are the simulation results. According to Section 3, the orientation errors of simulation results of the four typical approaches are almost exactly the same, so only one curve is drawn here to facilitate the observation and comparison of simulation and experimental results. In addition, the dithering of the experiment orientation error curves is due to cloud interference, which is not the focus of this paper and would not be discussed in detail here.

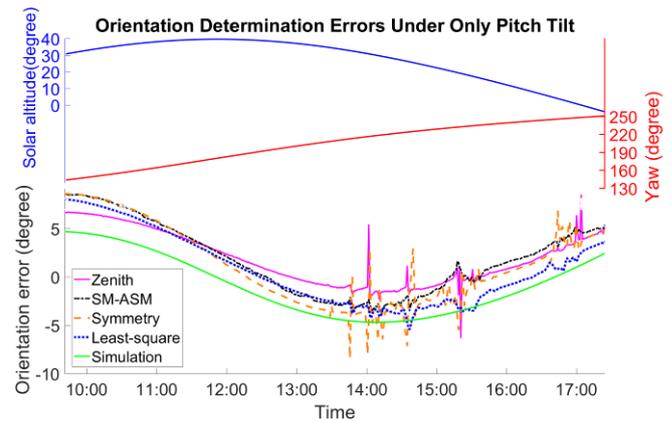

Fig.17 Field experiment under only pitch tilt condition on 15 November 2019, where the pitch angle is -20.0°.

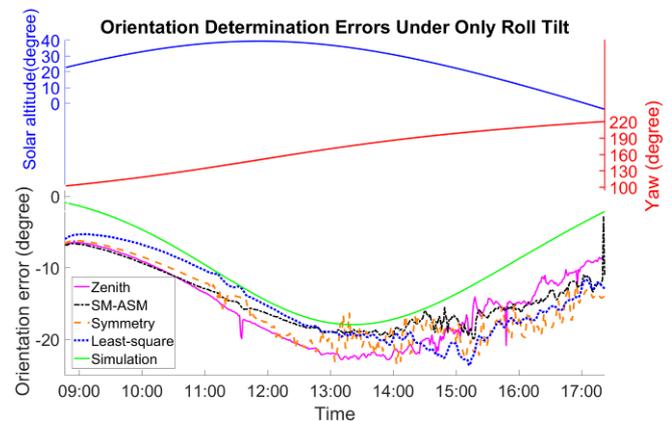

Fig.18 Field experiment under only roll tilt condition on 16 November 2019, where the roll angle is 29.1°..

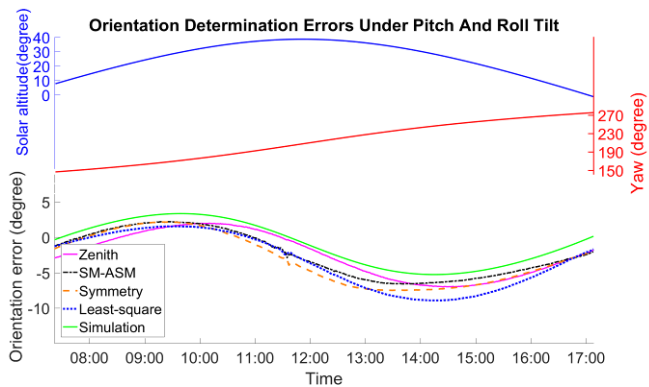

Fig.19 Field experiment under pitch and roll tilts on 19 November 2019, where the pitch angle is -16.3° and the roll angle is -9.9°.

As illustrated in Fig. 17, 18 and 19, comparing the simulation results with the experiment results of the four typical approaches, it is clear that: (Ⅰ) There are some differences of the simulation results and the field experimental results, which are always maintained within a range; (Ⅱ) The experiment error curves and simulation error curves have the same variation trend, and the experiment errors of the four typical approaches have the same variation trend compared with each other.

For further analysis, in (Ⅰ) of the above paragraph, Rayleigh sky model is an ideal model which only considers a single scattering event, and has some differences from the actual skylight polarization pattern [44]. So, the experiment error curves do not coincide with the simulation error curves.

For (Ⅱ) of the above paragraph, no matter only pitch tilt condition, only roll tilt condition, pitch and roll tilts condition, the orientation error curves of field experiments and simulation have the same variation, and the experiment errors of the four typical approaches have the same variation trend compared with each other. All these further showed that the orientation error characteristics of the four typical approaches are consistent under tilt interference. In simulation, the simulation error curves of the four typical approaches almost coincide. However, field experiment error curves of the four typical approaches do not coincide with each other. This is because field experiments have the influence of not only sensor tilts, but also some other disturbances, such as measurement noise and clouds. Especially for the part where the curve wobbles a lot, orientation determination is disturbed by the clouds.

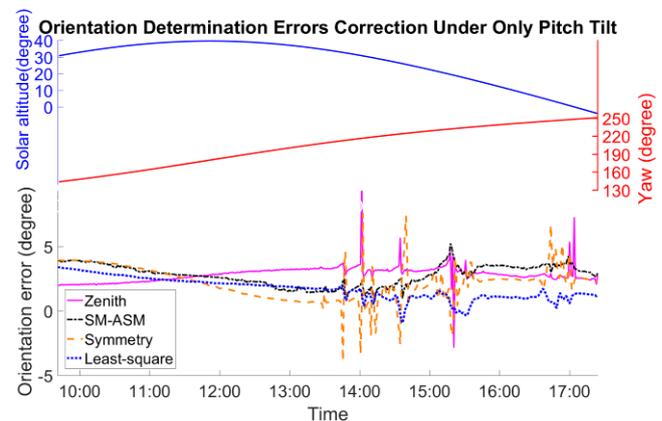

Fig.20 Orientation determination errors correction under only pitch tilt condition on 15 November 2019, where the pitch angle is -20.0°.

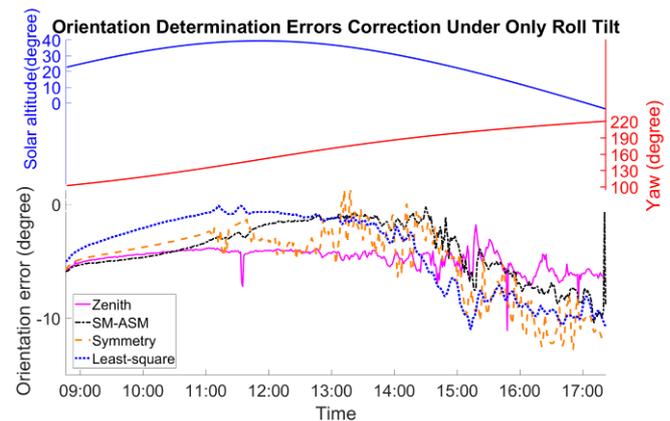

Fig.21 Orientation determination errors correction under only roll tilt condition on 16 November 2019, where the roll angle is 29.1°.

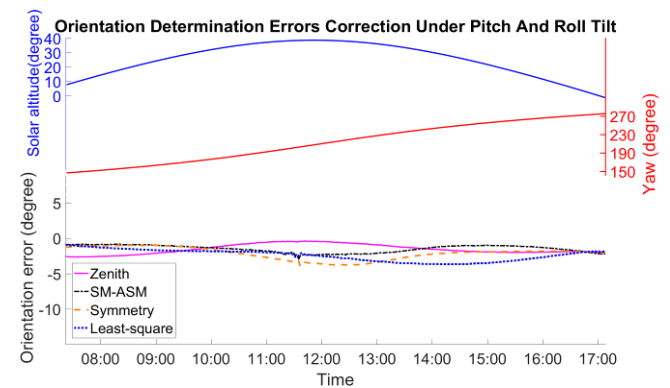

Fig.22 Orientation determination errors correction under pitch and roll tilts on 19 November 2019, where the pitch angle is -16.3° and the roll angle is -9.9°.

As illustrated in Fig. 20, 21 and 22, The results of this paper can be used to correct the orientation determination errors caused by sensor tilts. Especially, when the weather is fine, the error correction result is better, as shown in Fig. 22. However, when the weather conditions are quite complex and there is the interference of clouds, the results of this paper have some effect on error correction, but the effect is poor, as shown in Fig. 20, especially Fig. 21. This is because, when the weather is clear, skylight polarization patterns are closer to the Rayleigh sky model, and there is less interference from meteorological factors such as clouds. However, when the weather conditions become complicated, skylight polarization patterns deviate from Rayleigh sky model, and there is strong interference of meteorological factors such as clouds, resulting in poor correction effect.

## 4 Conclusion

In this study, the influence of sensor tilts on polarized skylight orientation determination is investigated in detail. Four typical polarization orientation determination approaches are described and compared with each other under only pitch tilt condition, only roll tilt condition, pitch and roll tilts condition. Simulation based on Rayleigh sky model shows that the error characteristics of the four approaches are completely consistent and the error curves almost coincide when only affected by sensor tilt. With the increase in tilt and solar altitude, the orientation errors of the four approaches all tend to increase, and the orientation errors are also affected by yaw angle. In addition, field experiments show that the errors of the four approaches have the same variation trend. All these provide an important reference for the practical application of polarization orientation determination, especially for the installation error of sensor, the tilt of application platform, the change of three-dimensional attitude of carrier and so on.

Moreover, the results of this paper can also mitigate orientation determination errors caused by sensor tilts, and estimate the allowable range of pitch and roll angles when given the allowable error of orientation determination.

Based on the simulations and experiments, the influence of sensor tilts is investigated in detail. However, polarization orientation determination can be influenced not only by the sensor tilt, but also the measurement noise and clouds, how to eliminate these impacts for orientation determination would be the focus of our future research.

**References**：

[1] MARTIN B, AXEL B, SILVèRE B. AI-IMU Dead-Reckoning [J]. ArXiv, 2019,

[2] LUCAS P V, CLAUDINE B, FERNANDO A C, et al. A single sensor system for mapping in GNSS-denied environments [J]. Cognitive Systems Research, 2019, 56(246-61.

[3] PHAM K D, CHEN G, AYCOCK T, et al. Passive optical sensing of atmospheric polarization for GPS denied operations [M]//KHANH D P, GENSHE C. Sensors and Systems for Space Applications IX. 2016.

[4] LIU G-X, SHI L-F. Adaptive algorithm of magnetic heading detection [J]. Measurement Science and Technology, 2017, 28(11):

[5] JINKUI CHU, LE GUAN, SHIQI LI, et al. Atmospheric polarization field pattern distribution and polarization navigation technology [J]. Journal of Remote Sensing, 2018, 22(6):

[6] ZHENG YOU, ZHAO K. Space situational awareness system based on bionic polarization feature sensing and navigation information fusion [J]. Journal of Remote Sensing, 2018, 22(6):

[7] WEHNER R. Desert ant navigation: how miniature brains solve complex tasks [J]. J Comp Physiol A Neuroethol Sens Neural Behav Physiol, 2003, 189(8): 579-88.

[8] WINTER C M, BREED M D. Homeward navigation in Pogonomyrmex occidentalis harvester ants [J]. Insectes Sociaux,

2016, 64(1): 55-64.

[9] EVANGELISTA C, KRAFT P, DACKE M, et al. Honeybee navigation: critically examining the role of the polarization compass [J]. Philos Trans R Soc Lond B Biol Sci, 2014, 369(1636): 20130037.

[10] KRAFT P, EVANGELISTA C, DACKE M, et al. Honeybee navigation: following routes using polarized-light cues [J]. Philos Trans R Soc Lond B Biol Sci, 2011, 366(1565): 703-8.

[11] MUHEIM R. Behavioural and physiological mechanisms of polarized light sensitivity in birds [J]. Philos Trans R Soc Lond B Biol Sci, 2011, 366(1565): 763-71.

[12] SCHMELING F, TEGTMEIER J, KINOSHITA M, et al. Photoreceptor projections and receptive fields in the dorsal rim area and main retina of the locust eye [J]. J Comp Physiol A Neuroethol Sens Neural Behav Physiol, 2015, 201(5): 427-40.

[13] HOMBERG U. Sky Compass Orientation in Desert Locusts-Evidence from Field and Laboratory Studies [J]. Front Behav Neurosci, 2015, 9(346.

[14] GáBOR H. Polarized Light and Polarization Vision in Animal Sciences (Second Edition) [M]. Berlin, Germany: Springer, 2014.

[15] GKANIAS E, RISSE B, MANGAN M, et al. From skylight input to behavioural output: A computational model of the insect polarised light compass [J]. PLoS computational biology, 2019, 15(7):

[16] PHAM K D, COX J L, AYCOCK T M, et al. Using Atmospheric Polarization Patterns for Azimuth Sensing [M]//KHANH D P. Sensors and Systems for Space Applications VII. 2014.

[17] BOHREN C F. Atmospheric Optics [M]. Pennsylvania Pennsylvania State University, 2007.

[18] ALEXANDER A K. Light Scattering Reviews 10 [M]. Berlin, Germany: Springer, 2016.

[19] CHENG H, CHU J, ZHANG R, et al. Underwater polarization patterns considering single Rayleigh scattering of water molecules [J]. International Journal of Remote Sensing, 2019, 41(13): 4947-62.

[20] ISTVáN P, GáBOR H, RüDIGER W. How the clear-sky angle of polarization pattern continues underneath clouds: full-sky measurements and implications for animal orientation [J]. The Journal of Experimental Biology, 2001, 204(

[21] BENCE S, GA´BOR H T. How well does the Rayleigh model describe the E-vector distribution of skylight in clear and cloudy conditions? A full-sky polarimetric study [J]. JOURNAL OF THE OPTICAL SOCIETY OF AMERICA A-OPTICS IMAGE SCIENCE AND VISION, 2004, 21(9):

[22] RAYLEIGH L. On the scattering of light by small particles [J]. Philosophical Magazine, 1871, 41(275): 447-54.

[23] DUPEYROUX J, SERRES J R, VIOLLET S. AntBot: A six-legged walking robot able to home like desert ants in outdoor environments [J]. Science Robotics, 2019, 4(0307): 12.

[24] DUPEYROUX J, VIOLLET S, SERRES J R. An ant-inspired celestial compass applied to autonomous outdoor robot navigation [J]. Robotics and Autonomous Systems, 2019, 117(40-56.

[25] DUPEYROUX J, VIOLLET S, SERRES J R. Polarized skylight-based heading measurements: a bio-inspired approach [J]. J R Soc Interface, 2019, 16(150): 20180878.

[26] LU H, ZHAO K, YOU Z, et al. Angle algorithm based on Hough transform for imaging polarization navigation sensor [J]. Opt Express, 2015, 23(6): 7248-62.

[27] TANG J, ZHANG N, LI D, et al. Novel robust skylight compass method based on full-sky polarization imaging under harsh conditions [J]. Opt Express, 2016, 24(14): 15834-44.

[28] LU H, ZHAO K, WANG X, et al. Real-time Imaging Orientation Determination System to Verify Imaging Polarization Navigation Algorithm [J]. Sensors (Basel), 2016, 16(2): 144.

[29] GUAN L, LI S, ZHAI L, et al. Study on skylight polarization patterns over the ocean for polarized light navigation application [J]. Appl Opt, 2018, 57(21): 6243-51.

[30] ZHANG W, CAO Y, ZHANG X, et al. Angle of sky light polarization derived from digital images of the sky under various conditions [J]. Appl Opt, 2017, 56(3): 587-95.

[31] TAO M, XIAOPING H, LILIAN Z, et al. An Evaluation of Skylight Polarization Patterns for Navigation [J]. Sensors, 2015, 15(19.

[32]    HUIJIE Z, WUJIAN X, YING Z, et al. Polarization patterns under different sky conditions and a navigation method based on the symmetry of the AOP map of skylight [J]. Optics Express, 2018, 26(22):

[33]    FAN C, HU X, HE X, et al. Integrated Polarized Skylight Sensor and MIMU With a Metric Map for Urban Ground Navigation [J]. IEEE Sensors Journal, 2018, 18(4): 1714-22.

[34]    WANG Y, ZHANG L, HE X, et al. Multicamera polarized vision for the orientation with the skylight polarization patterns [J]. Optical Engineering, 2018, 57(04):

[35]    WANG Y, CHU J, ZHANG R, et al. Orthogonal vector algorithm to obtain the solar vector using the single-scattering Rayleigh model [J]. Appl Opt, 2018, 57(4): 594-601.

[36]    WANG Y, CHU J, ZHANG R, et al. A Bio-Inspired Polarization Sensor with High Outdoor Accuracy and Central-Symmetry Calibration Method with Integrating Sphere [J]. Sensors, 2019, 19(16):

[37]    LIANG H, BAI H, LIU N, et al. Limitation of Rayleigh sky model for bio-inspired polarized skylight navigation in three-dimensional attitude determination [J]. Bioinspir Biomim, 2020,

[38]    LUTHER P, MAYUR P. Blind Hexapod Walking Over Uneven Terrain Using Only Local Feedback [M]. IEEE International Conference on Robotics and Biomimetics. Phuket, Thailand. 2011: 1603-8.

[39]    RUI S, KEIGO W, ISAKU N. Laser odometry taking account of the tilt on the laser sensor [M]. 2015 10th Asian Control Conference (ASCC). Kota Kinabalu. 2015: 1-4.

[40]    ROBERTO G. An algorithm for the computation of the solar position [J]. Solar Energy, 2008, 82(462-70.

[41]    GRENA R. Five new algorithms for the computation of sun position from 2010 to 2110 [J]. Solar Energy, 2012, 86(5): 1323-37.

[42]    BERNARD B. Improvement in solar declination computation [J]. Solar Energy, 1985, 35(4): 367-9.

[43]    REDA I, ANDREAS A. Solar position algorithm for solar radiation applications [J]. Solar Energy, 2004, 76(5): 577-89.

[44]    BRINES M L, GOULD J L. Skylight polarization patterns and animal orientation [J]. Journal of Experimental Biology, 1982, 96(1):